# Mapping of the 2+1 Dirac-Moshinsky Oscillator Coupled to an External Isospin Field onto Jaynes-Cummings Model

A. S.-F. Obada[1], M. M. A. Ahmed[1], M. Abu-Shady[2] and H. F. Habeba[2]

Department of Mathematics, Faculty of Science, Al-Azher University, Nassr City 11884, Egypt[1]

Department of Mathematics and Computer Science, Faculty of Science, Menoufia University, Shebin Elkom 32511, Egypt[2]

**Abstract**

In this paper, the 2+1 Dirac-Moshinsky oscillator ( 2+1 DMO ) coupled to an external isospin field is mapped onto the Jaynes-Cummings model ( JCM ), which describes the interaction between two two-level systems and a quantum single-mode field. The time-dependent wave function and the density matrix are obtained in the two cases. In the first case, the quantum number state is considered, while in the second case, the coherent state is considered as an initial state. The effect of both the detuning parameter and the coherence angle are studied on the entanglement and on the population inversion. It has been shown that the coherent state gives good description than the number state for the entanglement and the population inversion.

# 1 Introduction

The Dirac oscillator was suggested [1, 2] and reinvestigated where the linear term $im\omega c\beta\alpha.r$ is added to the relativistic momentum of the free-particle Dirac equation [3, 4].

The 1+1 Dirac-Moshinsky oscillator (1+1 DMO) has been exactly solved by using the theory of the non-relativistic harmonic oscillator [5,6]. The 2+1 DMO has been related to quantum optics via the JCM [7, 8].

In quantum optics, the JCM is composed of a single two-level particle interacting with a single quantized cavity mode of the electromagnetic field [9]. This model is an important model in quantum optics because, it is exactly solvable in the rotating wave approximation and experimentally realized [10]. It was found that, the JCM had statistical properties that are not in classical fields. Such as the degree of coherence, squeezing and the occurrence of the collapse and revival phenomenon in the evolution of the atomic inversion [11, 12].

The connection to the quantum optics problem not only makes the solvability more understandable in a wider setting but also allows to conceive quantum optics experiments that emulate this system.

Most exciting properties of the DMO is its connection to quantum optics [13, 14]. The dynamics of the 2+1 DMO was studied [15], it is found that an exact mapping of this quantum relativistic system onto the JCM is obtained. In Ref. [16], the 2+1 DMO in an external magnetic field has been studied and the connection between anti JCM with the DMO in magnetic field was established without making any limit on the strength of the magnetic field.

The 1+1 and 2+1 DMO have been mapped to the JCM and the dynamical features of a Dirac particle under the influence of the external field have been studied at vacuum [17].

In this article, the 2+1 DMO is mapped to the JCM, in which the external isospin field is included. The wave function is obtained in the two cases: the number state and the coherent state. It is to be mentioned that the previous attempts [17] concentrated on number state and did not resort to use the coherent state. In addition, we study the behaviour of the isospin field with the DMO by calculating the entanglement and the population inversion.

This article is arranged as follows: Sec. 2 is devoted to introduce a brief summary of the 2+1 DMO and the JCM. Sec. 3 is devoted to explain the mapping of the DMO onto the JCM. In Sec. 4, we study the 2+1 DMO in the presence of an external isospin field. The study of dynamical features of this model especially the entanglement and the population inversion is performed in Sec. 5. Finally, some concluding remarks are presented in Sec 6.

# 2 Basic equations and relations

## 2.1 The 2+1 Dirac-Moshinsky Oscillator

The DMO is introduced by Moshinsky and Szczepaniak [3] by adding the linear term $im\omega c\beta\alpha.r$ to the usual Dirac Hamiltonian for a free particle. In the non-relativistic limit it corresponds to the harmonic oscillator plus a spin-orbit coupling term. The DMO model in 2+1 dimensions takes the following form [17]

$$\begin{aligned} i\hbar\frac{\partial}{\partial t}\left|\psi\right\rangle &= [\sum_{j=1}^{2}c\alpha_{j}(p_{j}+im\omega\beta r_{j})+mc^{2}\beta]\left|\psi\right\rangle \\ &= \hat{H}^{2}\left|\psi\right\rangle, \end{aligned} \quad (2.1)$$

where $c$ is the speed of light, $m$ is the rest mass of the particle, $\alpha_{j}$, $\beta$ are the Dirac matrices in the standard representation, taken here as $\alpha_{1}=-\hat{\sigma}_{y}$, $\alpha_{2}=-\hat{\sigma}_{x}$, $\beta=\hat{\sigma}_{z}$ where the $\hat{\sigma}$'s are the pauli matrices and $\omega$ turns out to be the harmonic oscillator frequency. We note that the standard Dirac equation is recovered at $\omega=0$ [18].

## 2.2 The Jaynes-Cummings model

The JCM [9] is a theoretical model in quantum optics. It describes the system of a two-level particle interacting with one mode of the electromagnetic field.

The Hamiltonian in the interaction picture takes the following form

$$\hat{H}_{JC} = \Omega(\hat{\sigma}_{+}\hat{a}+\hat{\sigma}_{-}\hat{a}^{\dagger})+\delta\hat{\sigma}_{z}, \quad (2.2)$$

where $\Omega$ is the particle-field coupling constant, the operators $\hat{\sigma}_+$ and $\hat{\sigma}_-$ are the usual raising and lowering operators for the two-level system that satisfy the commutation relations $[\hat{\sigma}_z, \hat{\sigma}_\pm] = \pm 2\hat{\sigma}_\pm$ and $[\hat{\sigma}_+, \hat{\sigma}_-] = \hat{\sigma}_z$. $\hat{a}^\dagger$ ($\hat{a}$) is the Boson creation (annhilation) operator which satisfy $[\hat{a}, \hat{a}^\dagger] = 1$. $\delta$ stands for the detuning of the atomic transition frequency from the cavity mode.

# 3   Mapping of the DMO model onto the JCM

In this section, we derive an exact mapping of the 2+1 DMO onto the JCM.

By considering the spinor $|\psi\rangle = \begin{bmatrix} |\psi_1\rangle \\ |\psi_2\rangle \end{bmatrix}$, and $\hat{H}^2 |\psi\rangle = E |\psi\rangle$, Eq. (2.1) becomes a set of coupled equations as follows

$$(E - mc^2) |\psi_1\rangle = c(ip_x - p_y + m\omega x + im\omega y) |\psi_2\rangle \tag{3.1}$$

$$(E + mc^2) |\psi_2\rangle = c(-ip_x - p_y + m\omega x - im\omega y) |\psi_1\rangle. \tag{3.2}$$

In order to find the solutions, it is convenient to introduce the following chiral creation and annihilation operators [17]

$$\hat{a}_r = \frac{(\hat{a}_x + i\hat{a}_y)}{\sqrt{2}}, \hat{a}_r^\dagger = \frac{(\hat{a}_x^\dagger - i\hat{a}_y^\dagger)}{\sqrt{2}}, \tag{3.3}$$

$$\hat{a}_l = \frac{(\hat{a}_x - i\hat{a}_y)}{\sqrt{2}}, \hat{a}_l^\dagger = \frac{(\hat{a}_x^\dagger + i\hat{a}_y^\dagger)}{\sqrt{2}}, \tag{3.4}$$

where

$$\hat{a}_j = \sqrt{\frac{m\omega}{2\hbar}} r_j + \frac{i}{\sqrt{2m\omega\hbar}} p_j, \tag{3.5}$$

$$\hat{a}_j^\dagger = \sqrt{\frac{m\omega}{2\hbar}} r_j - \frac{i}{\sqrt{2m\omega\hbar}} p_j. \tag{3.6}$$

Eqs. (3.1) and (3.2) can be rewritten in the following form

$$E |\psi_1\rangle = mc^2 |\psi_1\rangle + \eta \hat{a}_r |\psi_2\rangle, \tag{3.7}$$

$$E |\psi_2\rangle = \eta \hat{a}_r^\dagger |\psi_1\rangle - mc^2 |\psi_2\rangle, \tag{3.8}$$

where

$$\eta = 2\sqrt{mc^2\omega\hbar}, \tag{3.9}$$

$\hat{H}^2$ can be written in the following form

$$\hat{H}^2 = \eta(\hat{\sigma}_+\hat{a}_r + \hat{\sigma}_-\hat{a}_r^\dagger) + mc^2\hat{\sigma}_z. \tag{3.10}$$

In Eq. (3.10), the 2+1 DMO maps onto JCM. One identifies $\eta \to \Omega$, $mc^2 \to \delta$, the isospin with the atomic system and the subspace of $\hat{a}_r$ with the cavity mode.

# 4   the 2+1 DMO coupled to an external isospin field

Now, we study the 2+1 DMO in the presence of isospin field (atomic system). The dynamics of the 2+1 DMO in the isospin field $\Phi$ is governed by the Hamiltonian

$$\tilde{H} = \hat{H}^{(2)} + \Phi, \tag{4.1}$$

where $\hat{H}^{(2)}$ given by Eq. (3.10) and $\Phi$ is the hermitean operator takes the following form [19]

$$\Phi = (A + \hat{\sigma}_z B)(\acute{\sigma}_+\hat{a}_r + \acute{\sigma}_-\hat{a}_r^\dagger + \gamma\acute{\sigma}_z), \tag{4.2}$$

where $\acute{\sigma}$'s are the vectors of Pauli matrices, its have the same commutation relations as $\hat{\sigma}$'s, therefore, the corresponding ladder operators are defined by

$$\acute{\sigma}_\pm = \frac{1}{2}(\acute{\sigma}_x \pm i\acute{\sigma}_y). \tag{4.3}$$

We take the simplest form of $\Phi$ (i.e. linear) as the following form

$$\Phi = \chi(\acute{\sigma}_+\hat{a}_r + \acute{\sigma}_-\hat{a}_r^\dagger) + \gamma\acute{\sigma}_z. \tag{4.4}$$

By using the Heisenberg equation of motion $\frac{d\hat{O}}{dt} = \frac{1}{i\hbar}[\hat{O}, \hat{H}]$, where $\hat{O}$ is any operator, for the operators $\hat{n}_r = \hat{a}_r^\dagger\hat{a}_r$, $\hat{\sigma}_z$ and $\acute{\sigma}_z$, we deduce the following constant of motion

$$I = \hat{n}_r + \frac{1}{2}(\hat{\sigma}_z + \acute{\sigma}_z). \tag{4.5}$$

# 5 The dynamical features of the 2+1 DMO model

In this section, we study the dynamical features of the 2+1 DMO under the influence of the external isospin field by invoking the connection of the coupled DMO with quantum optics systems.

The Hamiltonian (4.1) can be described in quantum optics as the interaction between two two-level particles (atoms) and electromagnetic field where $\hat{a}_r$ the ladder operator of the cavity and each isospin with an atom also $\eta$ and $\chi$ can be described as the coupling of each atom to the cavity isospin and $mc^2$ and $\gamma$ as the detuning of each transition level with the cavity mode.

We use two different cases:

(i) Initial entanglement atoms and number state, in this case, the initial state of the total model can be written as

$$|\psi(0)\rangle = (\cos(\theta)\left|-\dot{+}\right\rangle + \sin(\theta)\left|+\dot{-}\right\rangle)|0\rangle \tag{5.1}$$

and we take the quantum number $N = 0$ (vacuum state).

So the state vector at $t > 0$ takes the following form

$$|\psi(t)\rangle = C_1(t)\left|-\dot{-},1\right\rangle + C_2(t)\left|+\dot{-},0\right\rangle + C_3(t)\left|-\dot{+},0\right\rangle. \tag{5.2}$$

The coefficients $C_j(t)$, $(j = 1, 2, 3)$ can be obtained by solving the Schrödinger equation $i\hbar\frac{\partial}{\partial t}|\psi(t)\rangle = \tilde{H}|\psi(t)\rangle$ where $\tilde{H}$ is given by Eq. (4.1). Therefore, we have the following system of differential equations for $C_j(t)$

$$i\dot{C}_1(t) = \eta C_2(t) + \chi C_3(t) - (mc^2 + \gamma)C_1(t), \tag{5.3}$$

$$i\dot{C}_2(t) = \eta C_1(t) + (mc^2 - \gamma)C_2(t), \tag{5.4}$$

$$i\dot{C}_3(t) = \chi C_1(t) + (-mc^2 + \gamma)C_2(t). \tag{5.5}$$

After straightforward calculations, the time-dependent coefficients $C_j(t)$, $(j = 1, 2, 3)$ are obtained.

(ii) Initial entanglement atoms and coherent state, in this case, the initial state of the total system can be written as

$$|\psi(0)\rangle = (c_1\left|-\dot{-}\right\rangle + c_2\left|+\dot{-}\right\rangle + c_3\left|-\dot{+}\right\rangle + c_4\left|+\dot{+}\right\rangle)|\alpha\rangle, \tag{5.6}$$

where $c_i$, $(i = 1, 2, 3, 4)$ are constants satisfying the condition

$$\sum_{i=1}^{4} |c_i|^2 = 1, \tag{5.7}$$

we consider

$$c_1 = \cos(\theta)\cos(\phi), c_2 = \sin(\theta)\sin(\phi), \tag{5.8}$$

$$c_3 = \cos(\theta)\sin(\phi), c_4 = \sin(\theta)\cos(\phi), \tag{5.9}$$

where $|\alpha\rangle$ is the coherent state of the electromagnetic field, it takes the following form

$$|\alpha\rangle = \sum_{n=0}^{\infty} q_n |n\rangle, \tag{5.10}$$

with

$$q_n = \exp(\frac{-|\alpha|^2}{2})\frac{\alpha^n}{\sqrt{n!}}. \tag{5.11}$$

By using the constant of motion Eq. (4.5), the state vector $|\psi(t)\rangle$ at $t > 0$ takes the following form

$$|\psi(t)\rangle = \sum_{n=0}^{\infty}(B_1(n,t)\left|-\grave{-},n+2\right\rangle + B_2(n,t)\left|+\grave{-},n+1\right\rangle$$

$$+ B_3(n,t)\left|-\grave{+},n+1\right\rangle + B_4(n,t)\left|+\grave{+},n\right\rangle). \tag{5.12}$$

The coefficients $B_j(n,t)$, $(j = 1, 2, 3, 4)$ can be obtained by solving the Schrödinger equation. Therefore, we have the following system of differential equations for $B_j(n,t)$

$$i\dot{B}_1(n,t) = \lambda_2(n)B_2(n,t) + \lambda_4(n)B_3(n,t) - \zeta B_1(n,t), \tag{5.13}$$

$$i\dot{B}_2(n,t) = \lambda_2(n)B_1(n,t) + \lambda_3(n)B_4(n,t) + \xi B_2(n,t), \tag{5.14}$$

$$i\dot{B}_3(n,t) = \lambda_1(n)B_4(n,t) + \lambda_4(n)B_1(n,t) - \xi B_3(n,t), \tag{5.15}$$

$$i\dot{B}_4(n,t) = \lambda_1(n)B_3(n,t) + \lambda_3(n)B_2(n,t) + \zeta B_1(n,t), \tag{5.16}$$

where

$$\lambda_1(n) = \eta\sqrt{n+1}, \lambda_2 = \lambda_1(n+1), \tag{5.17}$$

$$\lambda_3(n) = \chi\sqrt{n+1},\ \lambda_4 = \lambda_3(n+1), \tag{5.18}$$

$$\zeta = mc^2 + \gamma,\ \xi = mc^2 - \gamma. \tag{5.19}$$

By using the Laplace transform, the time-dependent coefficients $B_j(n,t)$, ($j = 1, 2, 3, 4$) are obtained, by taking $\eta = \chi$ and $mc^2 = \gamma$ (the field strength equal to the rest mass energy, i.e identical particles), one can find simple explicit solutions for these coefficients.

Having obtained the explicit form of the time-dependent state, we are in the position to compute all quantities related to it, so our next task is to investigate the entanglement between the isospin field and the DMO, so we have to evaluate the reduced density matrix of the isospin field by taking a partial trace over the DMO degree of freedom.

We can write the reduced density matrix in the following form

$$\grave{\rho}(t) = Tr_{DMO}(|\psi(t)\rangle \langle \psi(t)|), \tag{5.20}$$

where $|\psi(t)\rangle$ is the time-dependent state vector. $\grave{\rho}(t)$ of the first case takes the following form

$$\grave{\rho}(t) = \rho_{ee}(t) \left|\grave{+}\right\rangle \left\langle\grave{+}\right| + \rho_{gg}(t) \left|\grave{-}\right\rangle \left\langle\grave{-}\right|, \tag{5.21}$$

where

$$\rho_{ee}(t) = |C_3(t)|^2, \tag{5.22}$$
$$\rho_{gg}(t) = |C_1(t)|^2 + |C_2(t)|^2, \tag{5.23}$$
$$\rho_{eg}(t) = 0 = \rho_{ge}(t). \tag{5.24}$$

Also, $\grave{\rho}(t)$ of the coherent state takes the following form

$$\begin{aligned}\grave{\rho}(t) &= \rho_{ee}(t) \left|\grave{+}\right\rangle \left\langle\grave{+}\right| + \rho_{gg}(t) \left|\grave{-}\right\rangle \left\langle\grave{-}\right| + \\ &\quad \rho_{eg}(t) \left|\grave{+}\right\rangle \left\langle\grave{-}\right| + \rho_{ge}(t) \left|\grave{-}\right\rangle \left\langle\grave{+}\right|,\end{aligned} \tag{5.25}$$

where

$$\rho_{ee}(t) = \sum_{n=0}^{\infty}(|B_3(n,t)|^2 + |B_4(n,t)|^2), \qquad (5.26)$$

$$\rho_{gg}(t) = \sum_{n=0}^{\infty}(|B_1(n,t)|^2 + |B_2(n,t)|^2), \qquad (5.27)$$

$$\rho_{eg}(t) = \sum_{n=0}^{\infty}(B_3(n+1,t)B_1^*(n,t) + B_4(n+1,t)B_2^*(n,t)) = \rho_{ge}^*(t). \qquad (5.28)$$

## 5.1 The entanglement

This section is devoted to a discussion of the entanglement between the isospin field and the DMO through the von Neumann entropy, where, the dynamics described by Hamiltonian (4.1) leads to an entanglement between the isospin field and the DMO. The von Neumann entropy is defined in quantum mechanics as a generalization of the classical Boltzmann entropy [20],

$$S = -Tr[\hat{\rho}\ln\hat{\rho}], \qquad (5.29)$$

where $\hat{\rho}$ is the density operator for a given quantum system, we set Boltzmann constant $K = 1$. For an initial pure state of the system, the entropy of the total system vanishes ($S = 0$), while if $\hat{\rho}$ describes a mixed state, then $S \neq 0$. We can either use the atomic entropy $S(t)$ to measure the amount of entanglement. The atomic entropy can be expressed in terms of the eigenvalues $\lambda_i(t)$, $(i = 1, 2, 3, 4)$ for the reduced atomic density operator as

$$S(t) = -\sum_{i=1}^{i=4} \lambda_i(t) \ln \lambda_i(t), \qquad (5.30)$$

for one particle (atom), the atomic entropy can be written as

$$S(t) = -\lambda_-(t)\ln\lambda_-(t) - \lambda_+(t)\ln\lambda_+(t), \qquad (5.31)$$

where $\lambda_\pm(t)$ are the eigenvalues of the reduced density matrix $\hat{\rho}(t)$ which can be easily evaluated as the following form

$$\lambda_\pm(t) = \frac{1}{2} \pm \frac{1}{2}\sqrt{\langle\hat{\sigma}_x(t)\rangle^2 + \langle\hat{\sigma}_y(t)\rangle^2 + \langle\hat{\sigma}_z(t)\rangle^2}, \qquad (5.32)$$

where

$$\langle \hat{\sigma}_x(t) \rangle = 2\Re[\rho_{eg}(t)], \qquad (5.33)$$

$$\langle \hat{\sigma}_y(t) \rangle = 2\Im[\rho_{eg}(t)], \qquad (5.34)$$

$$\langle \hat{\sigma}_z(t) \rangle = \rho_{ee}(t) - \rho_{gg}(t). \qquad (5.35)$$

Now, we turn our attention to examine numerically the dynamics of the von Neumann entropy for different values of $\gamma$ and $\theta$, where it is plotted against $\eta t$. For the first case, we take $\theta = \frac{\pi}{2}$ in **Fig.** (1), we see that, at $\gamma = 0$ (zero detuning), the von Neumann entropy reaches the maximum value ($\ln 2$) at regular time, see **Fig.** (1a). We increase the value of $\gamma$ in order to visualize the influence of the detuning parameter in the von Neumann entropy, we note that, the number of fluctuations decreases and the amplitude increases comparison to **Fig.** (1a), see **Figs.** (1b, 1c). The same effect has been observed in [17].

To visualize coherence angle effect see **Fig.** (2) where we use different values of $\theta$, we show that S(t) starts from *zero*, then it followed by a sequence of fluctuations, this means that this system begins by a pure (disentangled) state as mentioned in Eq. (5.1) and the maximum value of S(t) equal to $\ln 2$ (fully entanglement), see **Figs.** (2a) and (2c) at $\theta = 0$ and $\theta = \frac{\pi}{2}$. In **Fig.** (2b) at $\theta = \frac{\pi}{4}$, the S(t) starts from $\ln 2$ (mixed state).

For the second case, in **Fig.** (3), by increasing the value of detuning parameter from *zero* to 2 and 4, we note that, the number of fluctuations decrease and the interaction between the DMO and the field is almost entangled and never reaches to pure state. In **Fig.** (4), it is observed that, when $\theta = 0$ and $\theta = \frac{\pi}{2}$, S(t) start from zero (pure state) then it oscillated arround $\ln 2$, see **Figs.** (4a, 4c). In **Fig.** (4b) the system is in entanglement at all times. Finally, we observe that, the coherent state gives description better than the number state for the entanglement between the isospin field and the DMO, where the coherent state has better properties than the number state [21, 22].

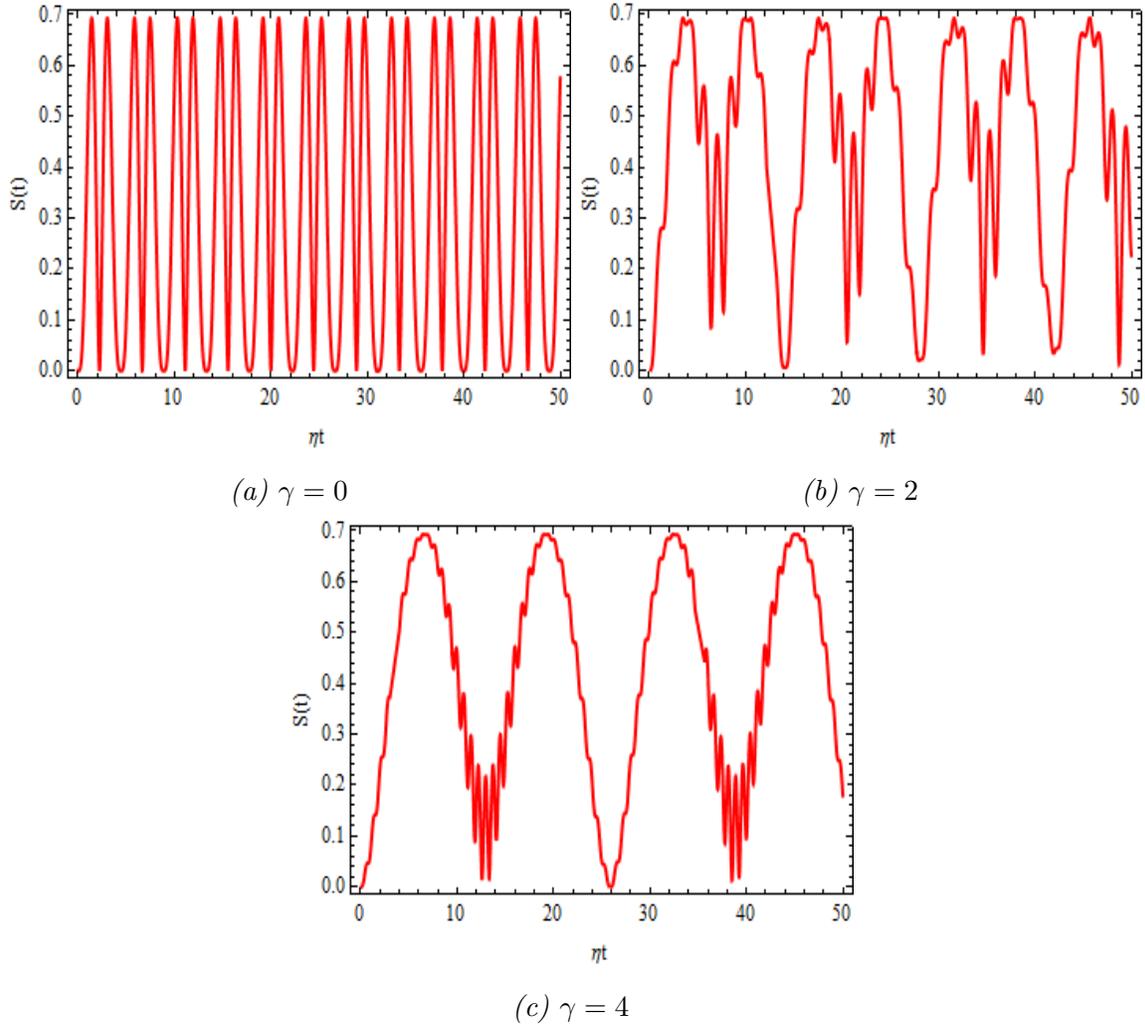

Figure 1: the von Neumann entropy is plotted as a function of $\eta t$ with $\theta = \frac{\pi}{2}$, $mc^2 = \gamma$ and $\eta = \chi$ ( the first case).

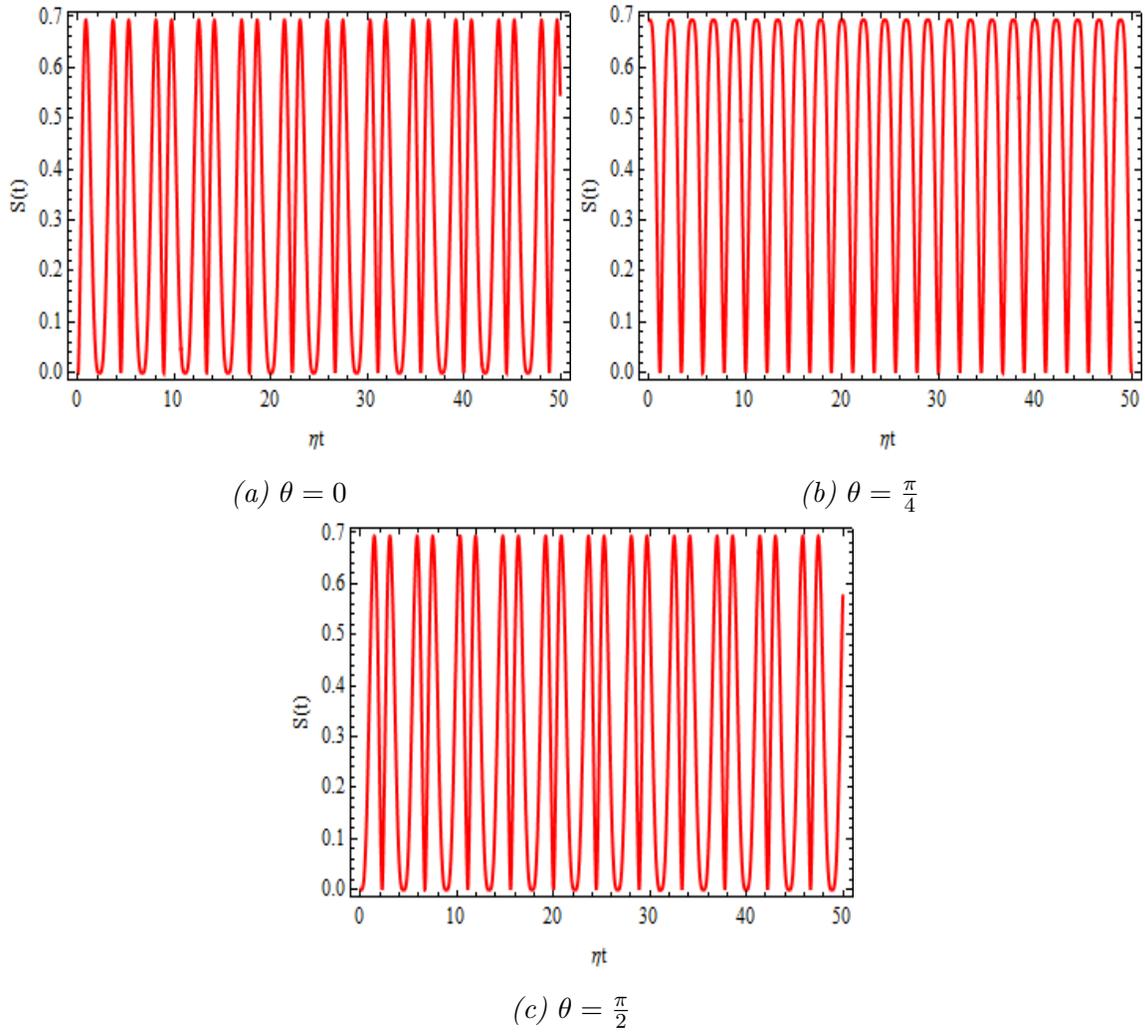

Figure 2: the von Neumann entropy is plotted as a function of $\eta t$ with $\gamma = 0$ and the other parameters are similar to **Fig.** (1) of the first case.

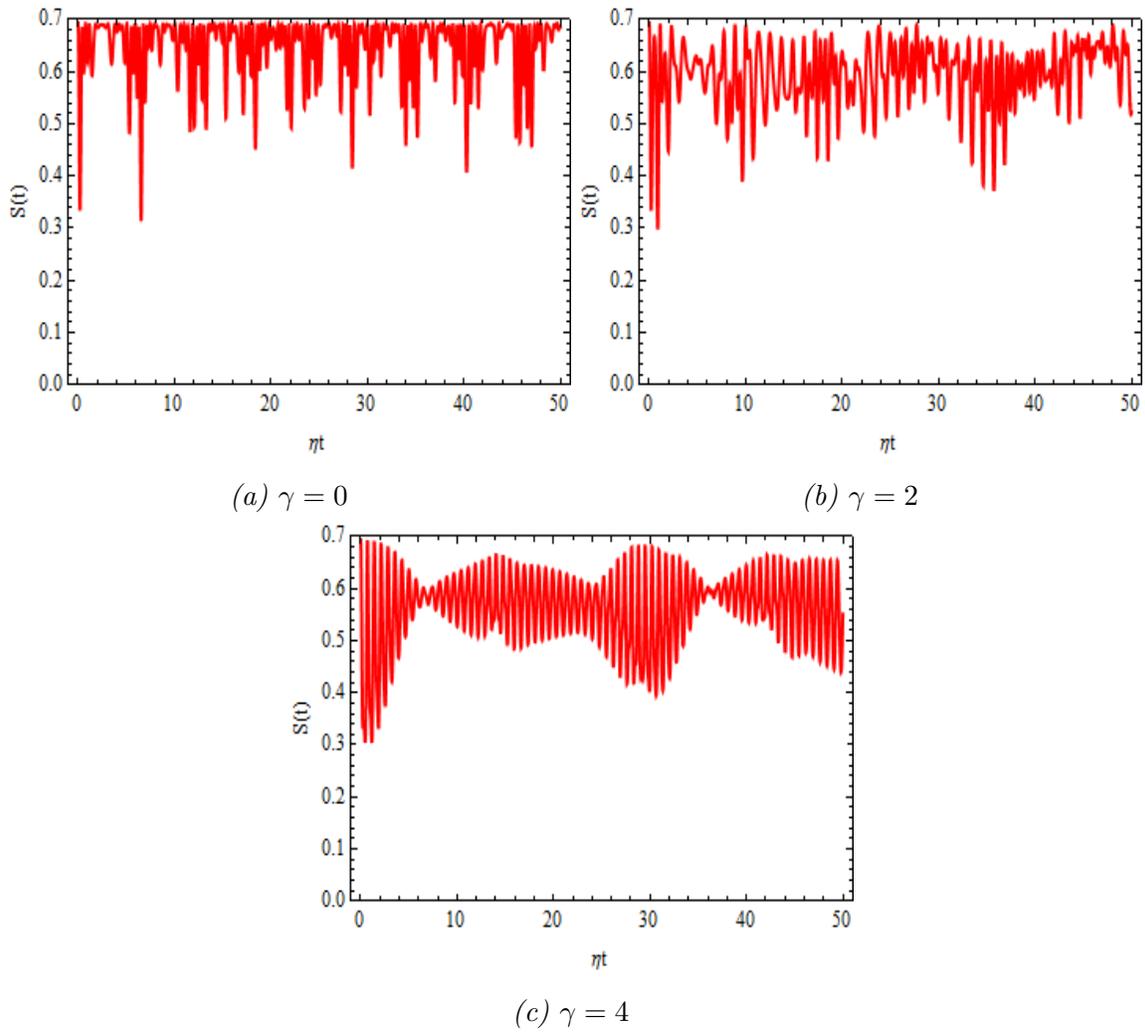

Figure 3: the von Neumann entropy is plotted as a function of $\eta t$ with $\theta = \frac{\pi}{4}$, $\phi = \frac{\pi}{2}$ and $\alpha = \sqrt{2}$ ( the second case).

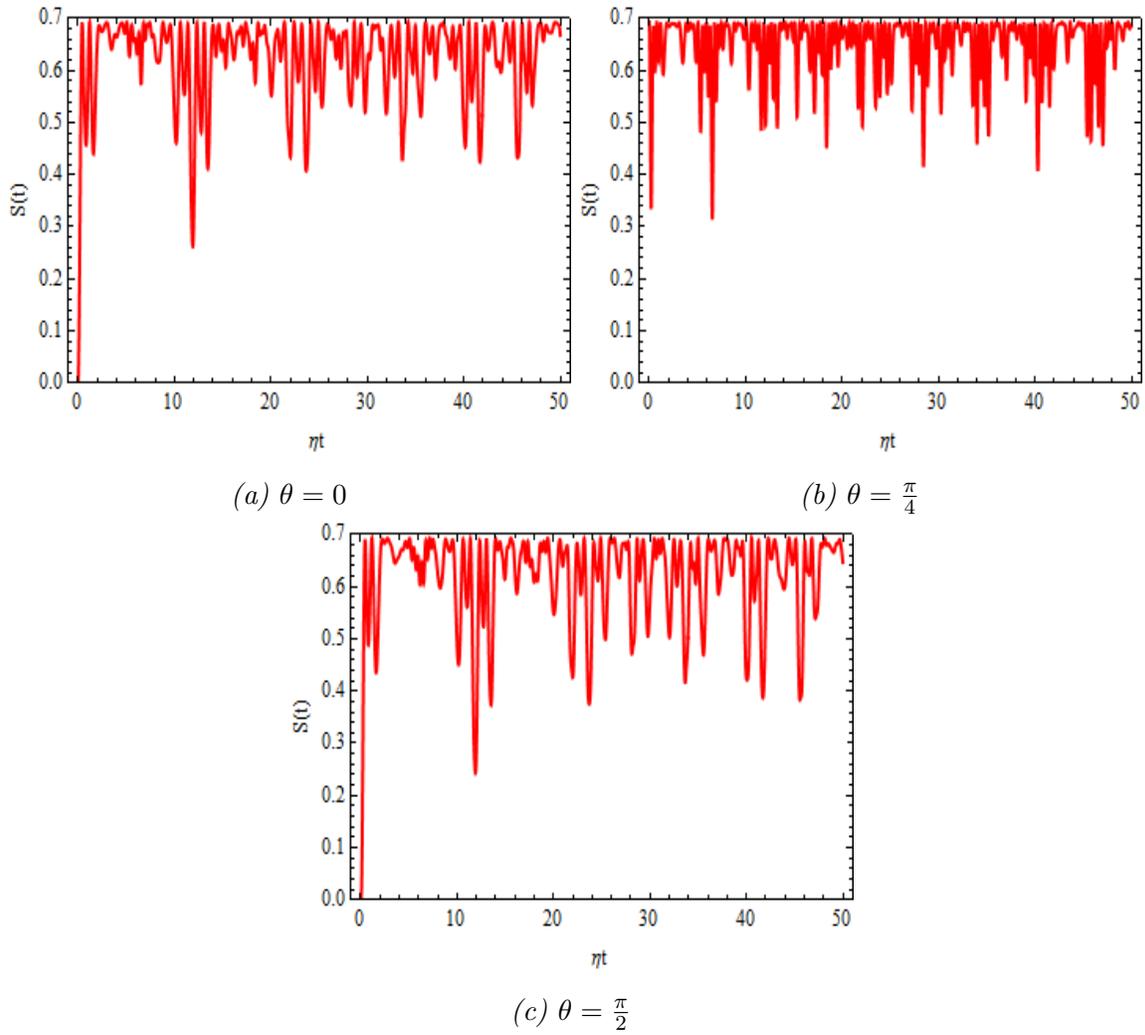

Figure 4: the von Neumann entropy is plotted as a function of $\eta t$ of the second case with $\gamma = 0$ and the other parameters are similar to **Fig.** (3).

## 5.2 The population inversion

The population inversion is defined as the difference between the probability of finding the isospin field in the excited state $|+\rangle$ and in the ground state $|-\rangle$. This in fact would give us information about the behaviour of the isospin field during the interaction period, which determines when this field reaches its maximal state and lead one to observe when this field is in its excited or ground state. From mathematical point of view the population inversion is the expectation value of the operator $\hat{\sigma}_z$, thus we have

$$W(t) = \rho_{ee}(t) - \rho_{gg}(t). \tag{5.36}$$

In **Figs.** (5, 7) the evolution of the population inversion with the scaled time $\eta t$ is displayed in order to see the effect of the detuning parameter in the behaviour of the isospin field in this model for the two cases and we study the effect of $\theta$ in **Figs.** (6, 8) where, we use the same initial parameters of the above figures ( the von Neumann entropy).

In **Fig.** (5), we note that the collapse and revival phenomenon not appear as expected in the number state and it is noted that there are regular fluctuation between excited and ground state of the isospin field (atom). By increasing the value of the detuning parameter, the amplitude of oscillations of the isospin field increases, see **Figs.**(5b, 5c). In **Fig.** (6), the effect of $\theta$ appears, it is observed that, the behavior of the isospin field at $\theta = 0$ is contrary to the behavior of this field at $\theta = \frac{\pi}{2}$, see **Figs.**(6a, 6c) where, at $\theta = 0$ the field oscillates from excited to ground state and the opposite appears at $\theta = \frac{\pi}{2}$ and there is regular fluctuation between excited and ground state. By taking $\theta = \frac{\pi}{4}$, **Fig.** (6b) observes that the field oscillates in the ground state and never reaches its excited state, also, the number of oscillations increase.

In the second case, the collapse and revival phenomenon appear clearly, which indicate that, the coherent state shows the behaviour of the isospin field during the interaction with DMO, which the number state failed to explain it, see **Fig.** (7). At $\gamma = 0$, the function $W(t)$ is symmetric around $W(t) = -0.06$, which means that, the isospin field oscillates in the ground state. By increasing the detuning parameter ($\gamma = 2$ or $\gamma = 4$) it is noted that, there are regular fluctuation between excited and ground state and the amplitude of oscillations increases as observed in **Figs.** (7b, 7c). The behaviour of the isospin field in **Fig.** (8) is similar to **Fig.** (6).

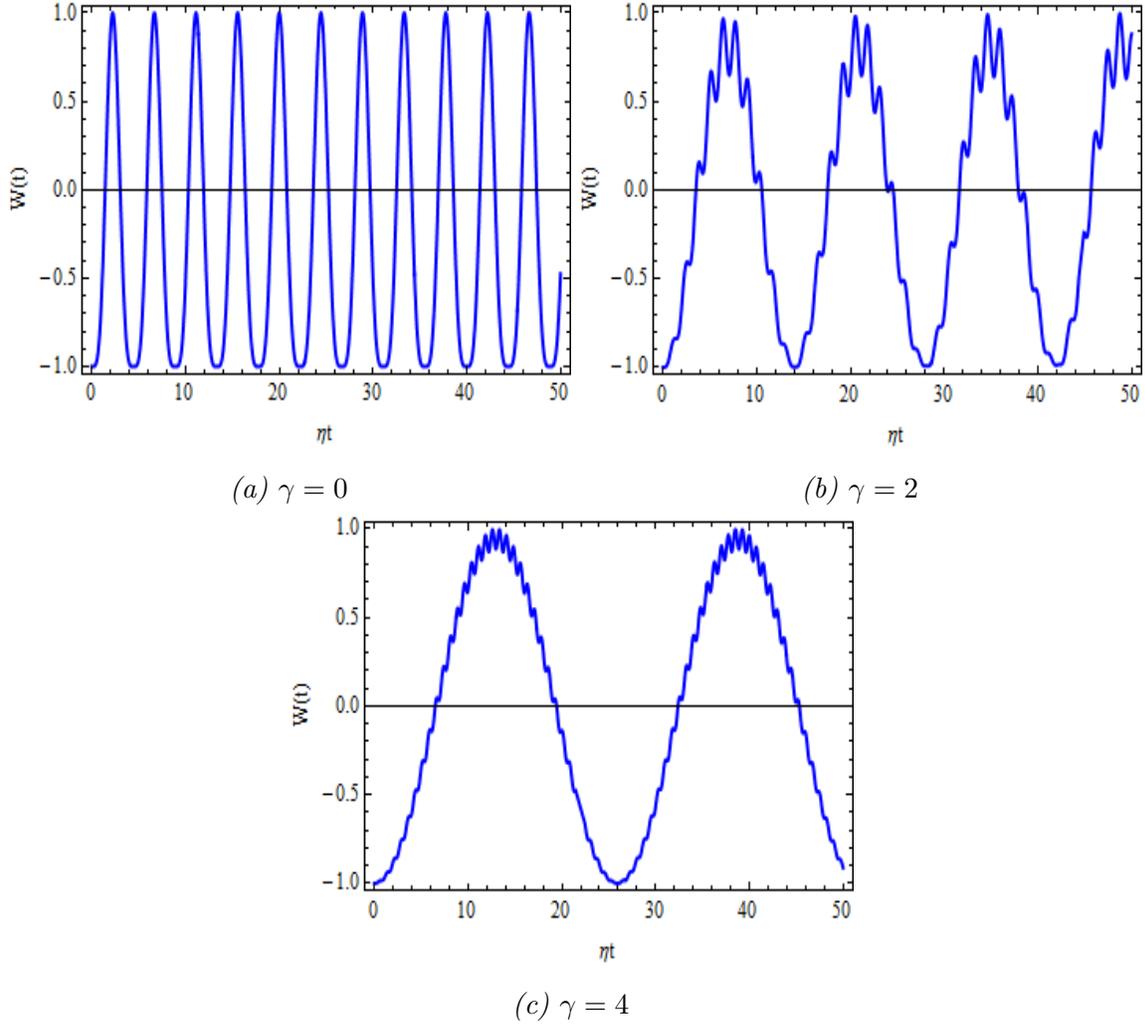

*Figure 5: the population inversion as a function of $\eta t$ of the first case. The parameters are similar to **Fig.** (1).*

The previous studies does not use the population inversion to show the behaviour of the isospin field during the interaction with DMO [15–17].

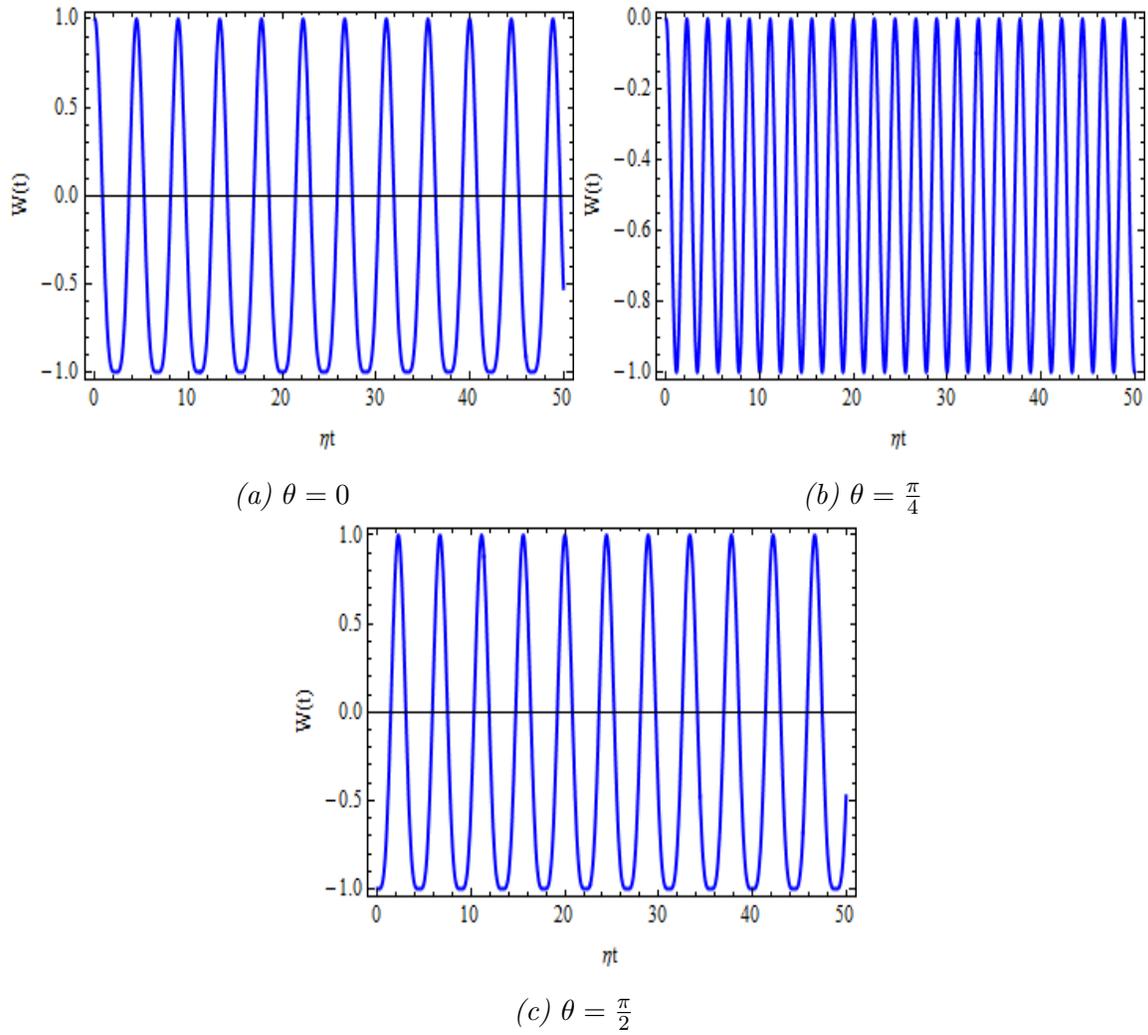

Figure 6: the population inversion as a function of $\eta t$ of the first case. The parameters are similar to **Fig.(** 2).

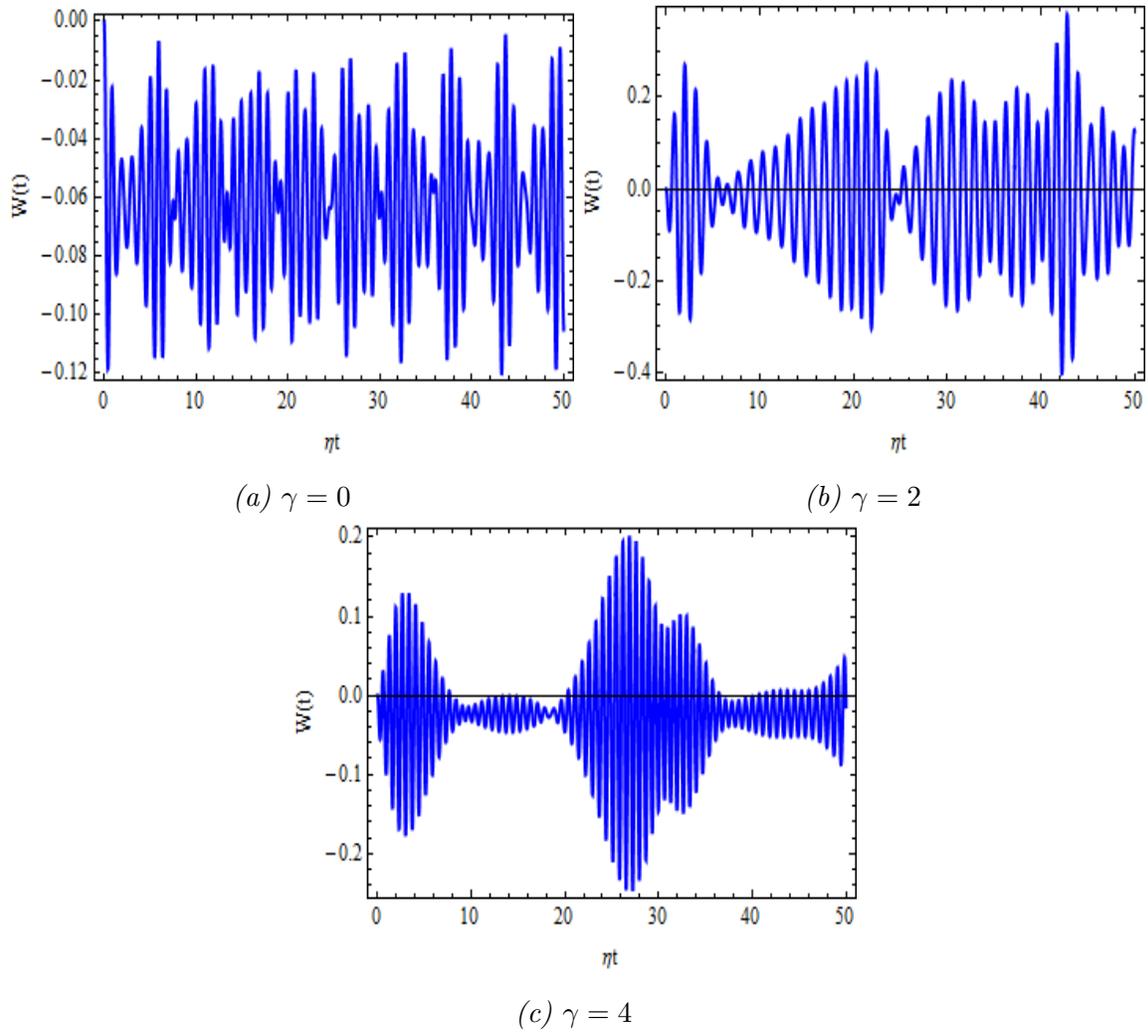

Figure 7: the population inversion as a function of ηt of the second case. The parameters are similar to **Fig.( 3)**.

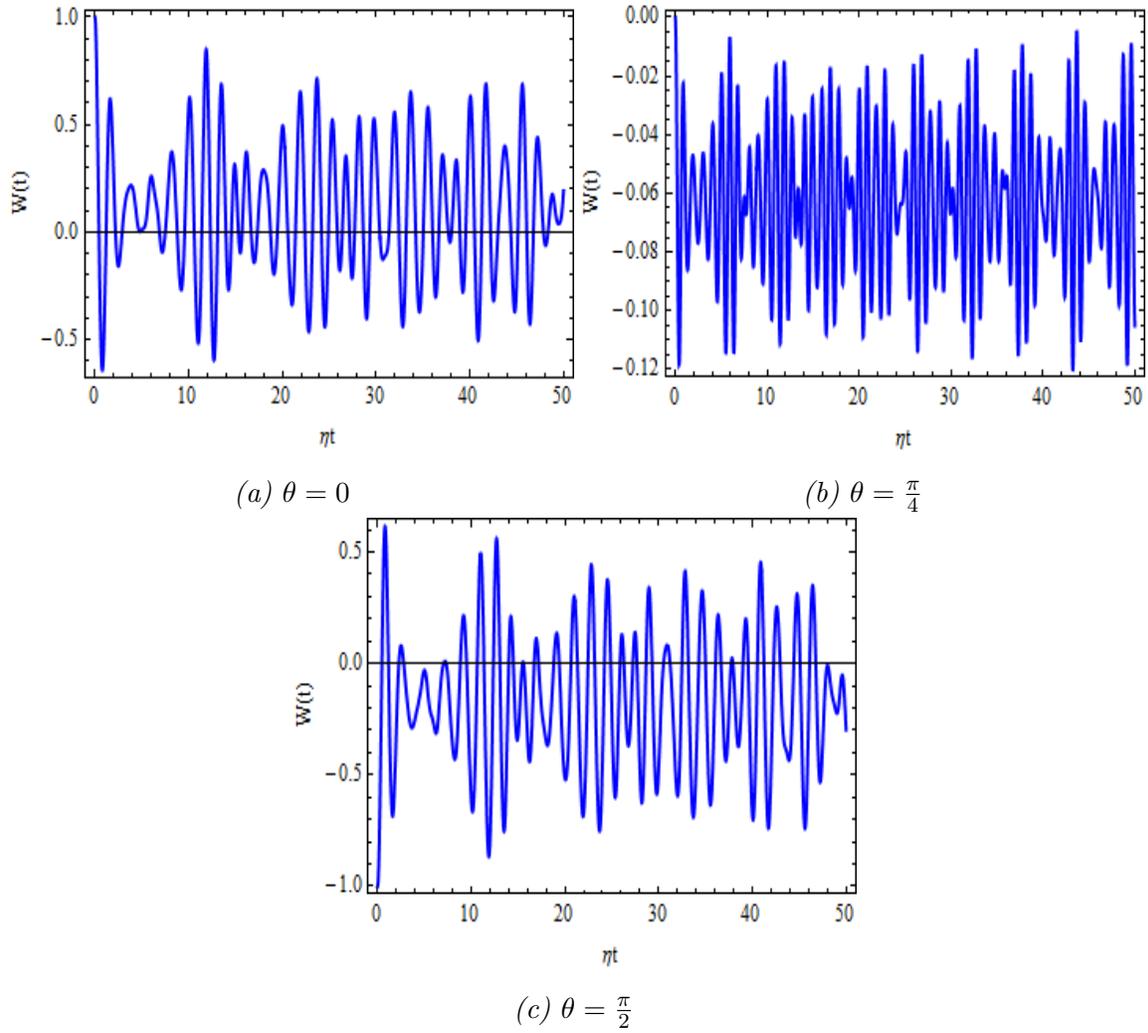

*Figure 8: the population inversion as a function of $\eta t$ of the second case. The parameters are similar to **Fig.( 4)**.*

# 6 Conclusion

We have studied how the 2+1 DMO coupled to an external isospin field can be mapped onto JCM. Also, we have studied the effect of $\gamma$ and $\theta$ on the entanglement and the population inversion. We have used two cases, in the first case, we have taken $N = 0$, and the second case, we have used coherent state as an initial state.

It is shown that, by increasing value of $\gamma$, the interaction between the isospin field and the DMO is almost entangled, the number of oscillations decreases and the amplitude of oscillations increases.

We conclude that the mapped DMO to JCM in the presence of external isospin field gives satisfied description of the entanglement and the population inversion in the coherent state so, this paper shows how important to link between the quantum optics and quantum relativistic.

# References

[1] D. Ito, K. Mori, and E. Carrieri, Nuovo Cimento A **51**, 119 (1967).

[2] P. A. Cook, Lett. Nuovo Cimento **1,** 419 (1971).

[3] M. Moshinsky and A. Szczepaniak, J. Phys. A **22**, L 817 (1989).

[4] D. Ojeda-Guillén, R. D. Mota, and V. D. Granados, J. Math. Phys. **57,** 062104 (2016).

[5] F. M. Toyama, Y. Nogami, and F. A. B. Coutinho, J. Phys. A: Math. Gen. **30**, 2585 (1997).

[6] R. Szmytkowski and M. Gruchowski, J. Phys. A: Math. Gen. **34**, 4991 (2001).

[7] A. Bermudez, M. A. Martin-Delgado, and E. Solano, Phys. Rev. A **76**, 041801 (2007).

[8] E. Sadurni, J. M. Torres, and T. H. Seligman, J. Phys. A: Math. Theor. **43**, 285204 (2010).

[9] E. T. Jaynes and F. W. Cummings, Proc. IEEE **51,** 89 (1963).

[10] S. Haroche and J. M. Raimond, Exploring the Quantum: Atoms, Cavities and Photons (Oxford University Press, Oxford, 2007).

[11] J. H. Eberly, N. B. Narozhny and J. J. S. -Mondragon, Phys. Rev. Lett., **44,** 1323 (1980).

[12] N. B. Narozhny, J. J. S. -Mondragon and J.H. Eberly, Phys. Rev. A, **23,** 236 (1981).

[13] A. Bermudez, M. A. Martin-Delgado and E. Solano, Phys. Rev. A, **76**, 041801 (2007).

[14] J. Benítez, R. P. Martinez y Romero, H. N. Núñez-Yépez and A. L. Salas-Brito, Phys. Rev. Lett. **64,** 1643 (1990); Erratum, Phys. Rev. Lett., **65,** 2085 (1990).

[15] D. P. Mandal and S. Verma, Phys. Lett. A, **374**, 1021 (2010).

[16] A. Bermudez, M. A. Martin-Delgado and E. Solano, Phys. Rev. A, **76,** 041801 (2007).

[17] J. M. Torres, E. Sadurmi and T. H. Seligman, AI P Conference Proceedings , **1323,** 301 (2010).

[18] W. Greiner, "Relativistic Quantum Mechanics: Wave Equations", (Springer, Berlin, 2000).

[19] E. Sadurni, J. M. Torres and T. H. Seligman, J. Phys. A, **43,** 285204 (2010).

[20] J. von Neumann, Mathematical Fundations of Quantum Mechanics, (Princaton University Press, Princaton 1955).

[21] M. O. Scully and M. S. Zubairy, Quantum Optics, (Cambridge University Press, 1997).

[22] C. C. Gerry and P. L. Knight, Introductory Quantum Optics, (Cambridge University Press, 2005).